\documentclass[a4paper,11pt]{article}
\usepackage{pos}

\title{LUXE: A new experiment to study non-perturbative QED in electron-laser and photon-laser collisions}

\author[1]{Yee Chinn Yap}

\affiliation[]{Deutsches Elektronen-Synchrotron DESY,\\
  Notkestr. 85, 22607 Hamburg, Germany}
\note{For the LUXE Collaboration.}

\emailAdd{yee.chinn.yap@desy.de}

\abstract{The LUXE experiment (Laser Und XFEL Experiment) is an experiment in planning at DESY Hamburg using the electron beam of the European XFEL. LUXE is intended to study collisions between a high-intensity optical laser pulse and 16.5 GeV electrons, as well as collisions between the laser pulse and high-energy secondary photons. This will elucidate quantum electrodynamics (QED) at the strong-field frontier, where the electromagnetic field of the laser is above the Schwinger limit. In this regime, QED is non-perturbative. This manifests itself in the creation of physical electron-positron pairs from the QED vacuum, similar to Hawking radiation from black holes. LUXE intends to measure the positron production rate in an unprecedented laser intensity regime. An overview of the LUXE experimental setup and its challenges and progress is given in this article, along with a discussion of the expected physics reach in the context of testing QED in the non-perturbative regime.} 

\FullConference{%
  41st International Conference on High Energy physics - ICHEP2022\\
  6-13 July, 2022\\
  Bologna, Italy
}

\begin{document}
\maketitle

\section{Introduction}

Laser Und XFEL Experiment (LUXE) \cite{CDR} is a proposed new experiment at DESY and European XFEL in Hamburg, Germany aiming to study non-perturbative quantum electrodynamics (QED) in electron-laser and photon-laser collisions. The high-power laser and the electron beam from XFEL collide at an angle of 17 degrees at a frequency of 1~Hz, determined by the laser frequency. The electron beam  has a frequency of 10 Hz, allowing 9 out of 10 of the electron bunches to be used for background studies. LUXE uses one electron bunch out of 2700 from the XFEL with each bunch containing $1.5\times 10^9$ electrons of 16.5 GeV. Two running modes are planned at LUXE: $e$-laser where the electron beam from XFEL is collided directly with the laser and $\gamma$-laser where the electron beam is first converted into photon beam via Bremsstrahlung with a target before colliding with the laser.

\section{Physics}

\begin{figure}[h!]
\centering
\includegraphics[width=.45\columnwidth]{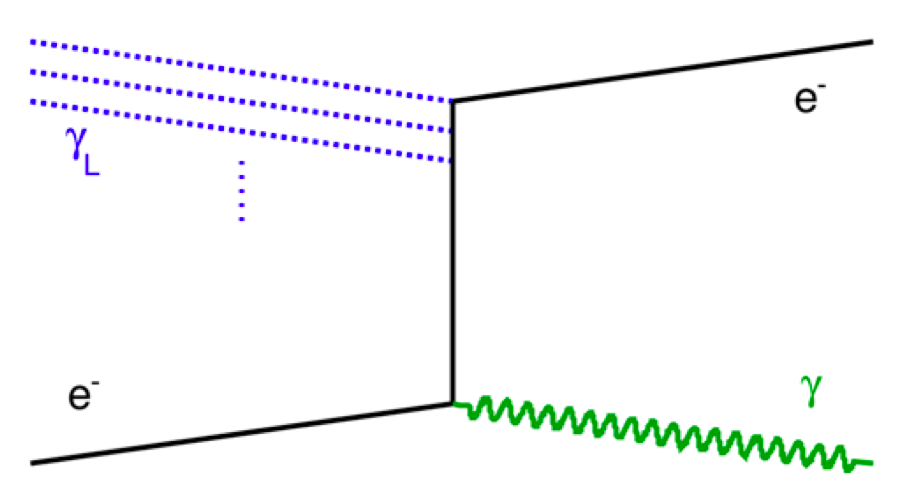}
\includegraphics[width=.45\columnwidth]{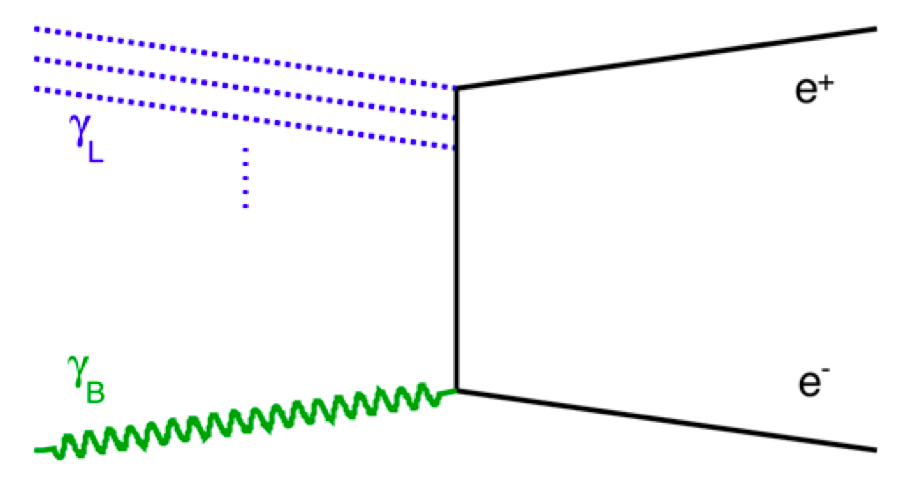}
\caption{Schematic diagrams for the non-linear Compton Scattering process and the non-linear Breit-Wheeler process.}
\label{fig:diagrams}
\end{figure}

Figure \ref{fig:diagrams} shows the two processes of interest at LUXE: the non-linear Compton scattering process of a photon radiated from the electron in the laser field,
\begin{equation}
  e^- + n \gamma_L \rightarrow e^- + \gamma,
\end{equation}
where $n$ is the number of laser photons $\gamma_L$ participating in the process; and the non-linear Breit-Wheeler pair creation,
\begin{equation}
  \gamma + n \gamma_L \rightarrow e^+ + e^-,
\end{equation}
from the interaction of a photon in the laser field. 

In the Breit-Wheeler process, the incoming photon can either be the Bremsstrahlung source photon in $\gamma$-laser mode or produced from the first Compton process in the $e$-laser mode. In $e$-laser interaction, both processes can occur either in two steps or in a single step ($e^- + n \gamma_L \rightarrow e^-e^+e^-$), collectively known as the non-linear trident process. The planned $\gamma$-laser mode allows real photon interactions with laser photons to be studied, as opposed to the two-step trident process.

An important parameter that characterises these interactions is $\xi$, the laser field intensity parameter or the charge-field coupling, defined as
\begin{equation}
  \xi=\frac{m_e E_L}{\omega_L E_{cr}},
  \label{eq:xi}
\end{equation}
where $m_e$ is the electron mass, $E_L$ is the laser field strength, $\omega_L$ is the frequency of the laser and $E_{cr}$ is the critical field strength, also known as the Schwinger limit defined as $E_{cr}=m_e^2c^3/e\hbar$.

Another useful parameter is the quantum non-linearity parameter $\chi$ defined as

\begin{equation}
\xi_i = \frac{\epsilon_i}{m_e} \frac{E_L}{E_{cr}}(1+\beta \cos \theta), i=e,\gamma
\end{equation}
where $\epsilon$ is the particle (electron or photon) energy, $\theta$ the collision angle and $\beta$ the speed in units of $c$. 

The processes are perturbative at low values of $\xi$ with the probability for the process involving $n$ laser photons given by $\xi^{2n}$ for $\xi \ll 1$. At larger $\xi$, one needs to consider contributions from all orders. LUXE aims to make precise measurements of the interactions in a transition from the perturbative to the non-perturbative regime.

\begin{figure}[h!]
\centering
\includegraphics[width=.7\columnwidth]{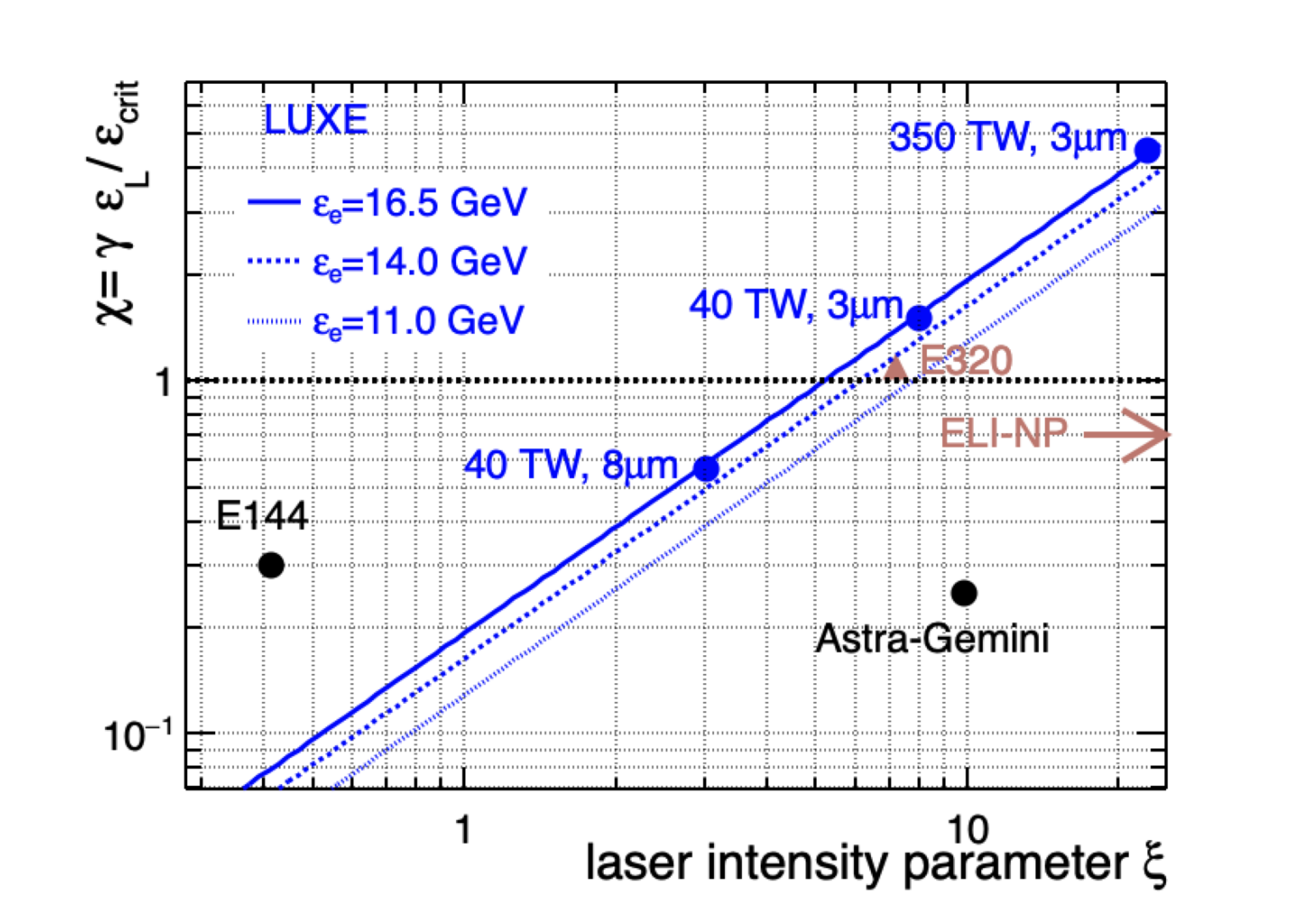}
\caption{$\chi$ vs $\xi$ for a selection of experiments and facilities. For LUXE, three beam energies are shown as isolines, and two laser focus spot sizes are highlighted for the phase-0 (40~TW) laser and one for the phase-1 (350~TW) laser. Reproduced from Ref.~\cite{CDR}.}
\label{fig:paramspace}
\end{figure}

Figure \ref{fig:paramspace} shows the QED parameter space of $\chi$ vs $\xi$ and the reach of several ongoing or planned experiments including LUXE. LUXE is planned in two phases: the first phase uses a $40$~TW laser, while the second phase uses an upgraded laser. LUXE spans a range of $\xi$ and $\chi$ parameters, from the perturbative to the non-perturbative region. E144 is an experiment at SLAC in the 1990s that reached $\chi=0.25$ and $\xi=0.4$, still within the perturbative regime, but already with observable non-linear effects where the trident process was observed. Other proposed experiments are E320 at SLAC and ELI-NP in Romania, while Astra-Gemini in the UK is ongoing. For a recent review of strong-field QED physics, see Ref. \cite{review}.

The abundant photons produced in LUXE, mainly through the non-linear Compton scattering, can be used to search for physics beyond the Standard Model (BSM). A scenario considered is the creation of Axion-like particles (ALPs) produced in the LUXE photon dump via the Primakoff effect. The ALPs would then decay into two photons and can be detected via dedicated detectors. Ref. \cite{BSM} studies the sensitivity of such a search. 

\section{Measurements}

The following measurements as a function of $\xi$ are planned. 

\begin{figure}[h!]
\centering
\includegraphics[width=.48\columnwidth]{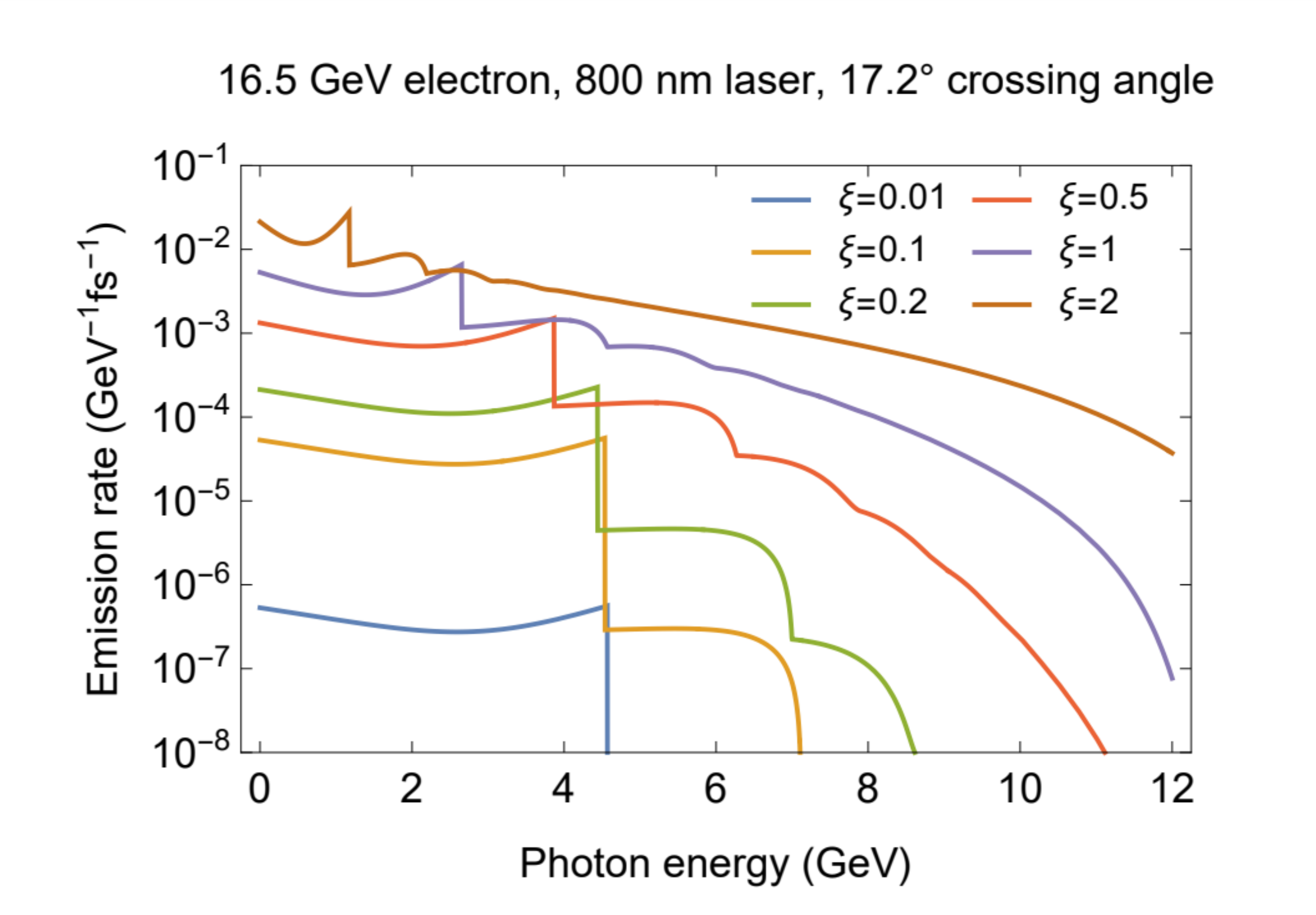}
\includegraphics[width=.48\columnwidth]{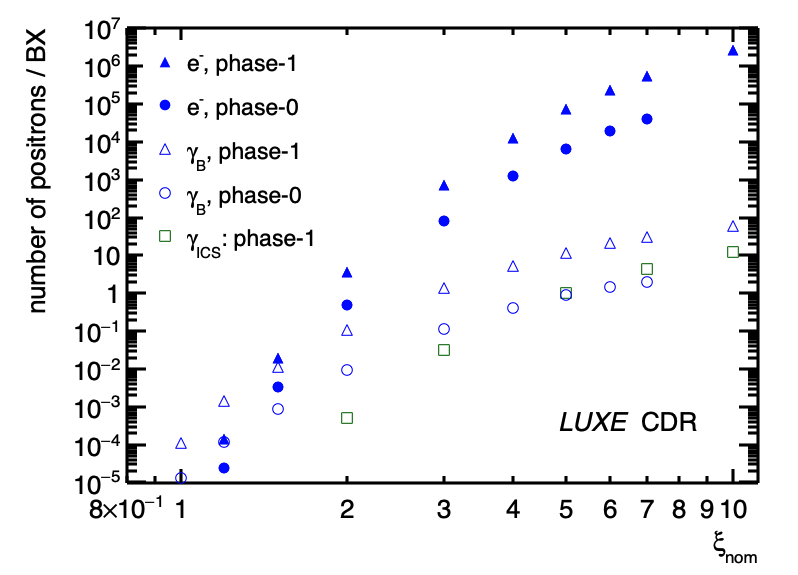}
\caption{(Left) Photon emission rate as a function of the photon energy for different values of $\xi$. (Right) Positron production rate as a function of $\xi$ for $e$-laser and $\gamma$-laser runs in phase-0 and phase-1. Reproduced from Ref.~\cite{CDR}.}
\label{fig:measurements}
\end{figure}

\begin{itemize}
\item The position of Compton edge determined from the electron and photon energy spectra in $e$-laser runs.  Figure \ref{fig:measurements} shows the photon energy spectrum for different $\xi$ values, and the edge is given by the first kink in the spectrum and it is seen to shift as a function of $\xi$ due to the effective mass gained by the electron $m_*=m_e \sqrt{1+\xi^2}$.
\item The positron rate which spans many orders of magnitudes, as shown in the right figure in Figure \ref{fig:measurements}. The positron production rate is equal to the rate of Breit-Wheeler process.
\item The number of photons radiated per electron in $e$-laser mode.
\end{itemize}

\section{Laser}

A Titanium Sapphire laser based on chirped pulse amplification technology is planned. The photon wavelength is $800$~nm, corresponding to $1.55$~eV in energy. Different $\xi$ values can be reached by changing the focus of the laser. Exceptional shot-to-shot stability in the laser intensity is needed in LUXE. 

Table \ref{tab:laser} outlines a few quantities related to the laser system in phase-0 and phase-1. The higher laser power in phase-1 allows higher values of $\xi$ and $\chi$ to be reached. The variation of these parameters is achieved by changing the laser focal spot waist, where the highest intensity is obtained by focussing the laser to $3$~\text{$\mu$m}.

\begin{table}[h!]
\centering
    \begin{tabular}{lll}
      \hline
      & Phase-0 & Phase-1 \\
      \hline
      Laser power (TW) & 40 & 350 \\
      Peak intensity in focus ($\times 10^{20} \text{W/cm}^2$) & <1.33 & <12 \\
      Dimensionless peak intensity $\xi$ & <7.9 & <23.6 \\
      Quantum parameter $\chi_e$ for $\epsilon_e=16.5$~GeV & <1.5 & < 4.45 \\
      Laser focal spot waist (\text{$\mu$m}) & \multicolumn{2}{c}{$\geq 3$} \\
      Laser pulse duration (fs) & \multicolumn{2}{c}{30} \\
      \hline
    \end{tabular}
  \caption{Parameters for the laser system as planned for the two phases of the LUXE experiment.}
  \label{tab:laser}
\end{table}

\section{Experimental setup}

\begin{figure}[h!]
\centering
\includegraphics[width=.48\columnwidth]{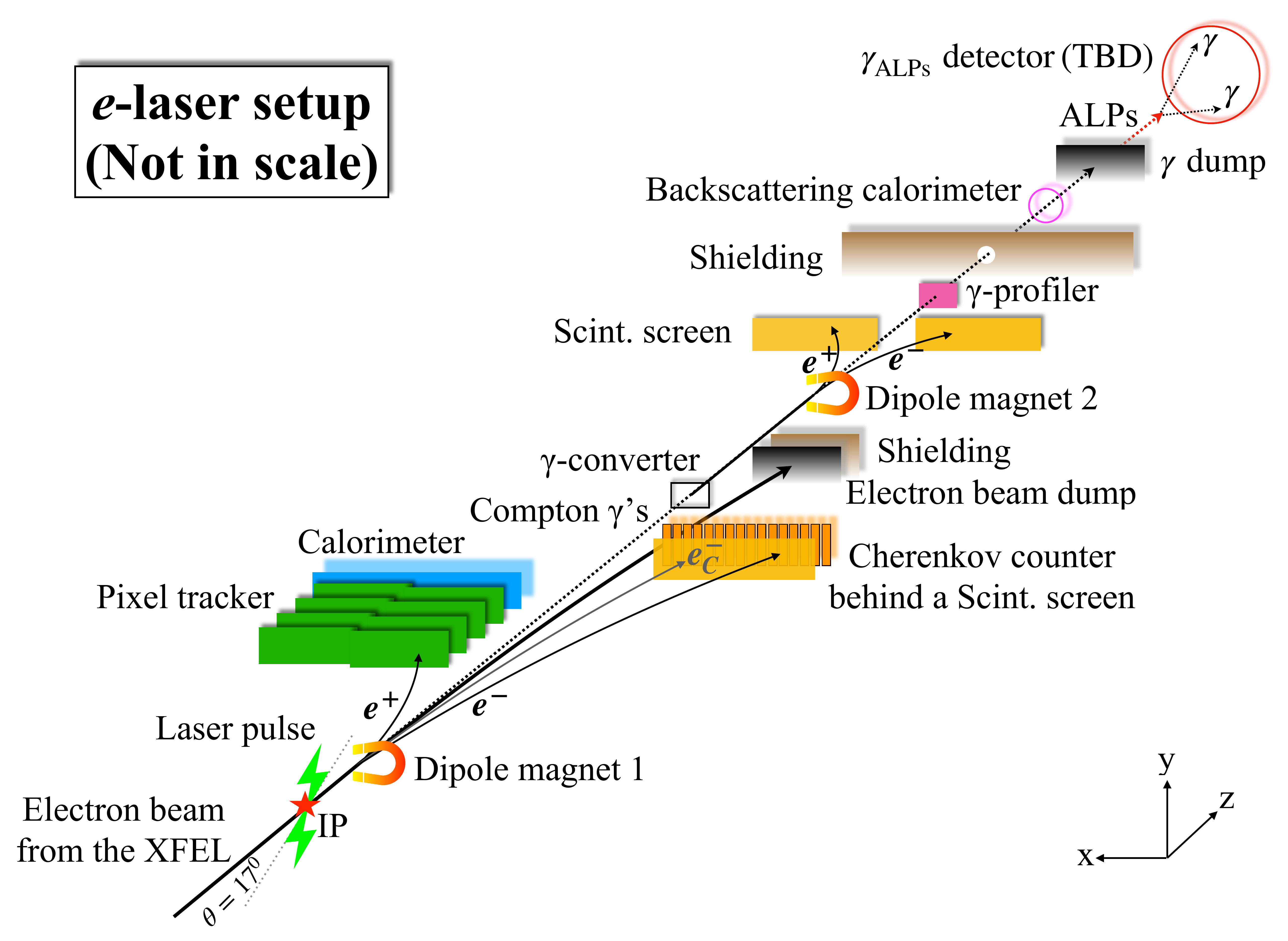}
\includegraphics[width=.48\columnwidth]{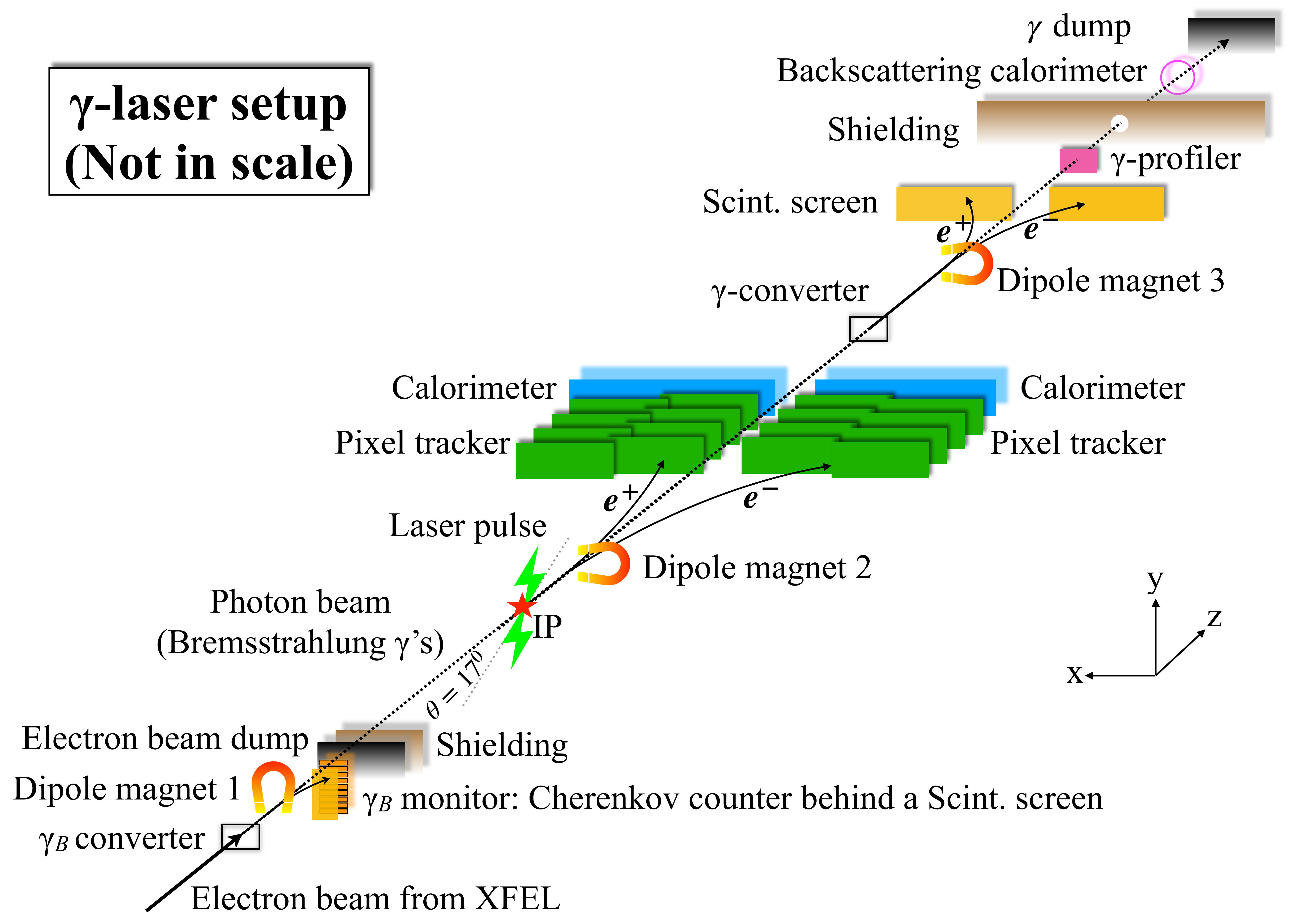}
\caption{Schematic layouts for the $e$-laser and $\gamma$-laser setup. Reproduced from Ref.~\cite{CDR}.}
\label{fig:detectors}
\end{figure}

Figure \ref{fig:detectors} shows the layout of the LUXE detectors in the two setups. 
In the $e$-laser setup, a dipole magnet is placed after the interaction point to separate the particles. Electrons and positrons are deflected in opposite directions and measured in dedicated detectors, while the photons travel straight. The expected fluxes of these particles vary, e.g. the electron flux is expected to be around $10^9$ while the number of positrons ranges from $10^{-3}$ to $10^6$. The electron and positron detection systems use different technologies due to the difference in expected flux. The positron side uses a precision tracker and a calorimeter while the electron side uses a scintillation screen and a Cherenkov detector. The photon detection system is downstream of the electron and positron detection systems, where a target converts the photons into electron-positron pairs before they are measured with scintillation screens, a gamma profiler and a gamma flux monitor. A BSM detector can be placed at the end after the photon dump. 

In the $\gamma$-laser setup, after the electron beam hits the converter target, a dipole magnet is placed to deflect the electron beam into a dump while the photon beam travels on to interact with the laser. A dipole magnet is again placed after the interaction point to split the electron-positron pairs as done for the $e$-laser case. In this setup, the expected electron and position fluxes are the same and are much lower than in the $e$-laser case. Hence, the electron detection system uses also a tracking system similar to the positron detector.

\section{Conclusions}
In summary, LUXE will study strong-field QED in an unprecedented regime using a high-intensity optical laser pulse and high-energy electrons from the XFEL electron beam. The high photon flux in LUXE can also be used in a BSM physics programme with competitive sensitivity to other experiments. LUXE has passed the stage-0 critical approval from the DESY management. The start of data taking is planned for 2026.

\acknowledgments

We thank the DESY technical staff for continuous assistance and the DESY directorate for their strong support
and the hospitality they extend to the non-DESY members of the collaboration. This work has benefited from
computing services provided by the German National Analysis Facility (NAF) and the Swedish National Infrastructure for Computing (SNIC).

\end{document}